\documentclass[%
aps,
pra,
reprint,
nofootinbib,
superscriptaddress,
amsmath,amssymb,
nofootinbib,
]{revtex4-1}

\usepackage{graphicx}
\usepackage{xcolor}
\usepackage{dcolumn}
\usepackage{bm}
\usepackage{physics}
\newcommand{\nts}{\ensuremath{\Delta{}t}}

\begin{document}

\title{Optimal Control for the Quantum Simulation of Nuclear Dynamics}

\author{Eric T. Holland}
\thanks{these authors contributed equally}
\affiliation{Lawrence Livermore National Laboratory, P.O. Box 808, L-414, Livermore, California 94551, USA}

\author{Kyle A. Wendt}
\thanks{these authors contributed equally}
\email{wendt6@llnl.gov}
\affiliation{Lawrence Livermore National Laboratory, P.O. Box 808, L-414, Livermore, California 94551, USA}

\author{Konstantinos Kravvaris}
\affiliation{Lawrence Livermore National Laboratory, P.O. Box 808, L-414, Livermore, California 94551, USA}

\author{Xian Wu}
\affiliation{Lawrence Livermore National Laboratory, P.O. Box 808, L-414, Livermore, California 94551, USA}

\author{W. Erich Ormand}
\affiliation{Lawrence Livermore National Laboratory, P.O. Box 808, L-414, Livermore, California 94551, USA}

\author{Jonathan L DuBois}
\affiliation{Lawrence Livermore National Laboratory, P.O. Box 808, L-414, Livermore, California 94551, USA}

\author{Sofia Quaglioni}
\affiliation{Lawrence Livermore National Laboratory, P.O. Box 808, L-414, Livermore, California 94551, USA}

\author{Francesco Pederiva}
\affiliation{Physics Department, University of Trento, Via Sommarive 14, I-38123 Trento, Italy}
\affiliation{INFN-TIFPA Trento Institute of Fundamental Physics and Applications, Via Sommarive, 14, I-38123 Trento, Italy}

\date{\today}

\begin{abstract}

We propose a method for enacting the unitary time propagation of two interacting neutrons at leading order of chiral effective field theory by efficiently encoding the nuclear dynamics into a single multi-level quantum device.  The emulated output of the quantum simulation shows that, by applying a single gate that draws on the underlying characteristics of the device, it is possible to observe multiple cycles of the nucleons' dynamics before the onset of decoherence.  Owing to the signal's longevity, we can then extract spectroscopic properties of the simulated nuclear system.  This allows us to validate the encoding of the nuclear Hamiltonian and the robustness of the simulation in the presence of quantum-hardware noise by comparing the extracted spectroscopic information to exact calculations.  This work paves the way for transformative calculations of dynamical properties of nuclei on near-term quantum devices.

\end{abstract}
\maketitle

\section{\label{sec:intro}Introduction}
First proposed in the 1980's by Feynman~\cite{feynman1982}, quantum computers have been proven to be exponentially more efficient than any classical algorithm for the simulation of many-particle systems that are described by non-relativistic quantum mechanics \cite{lloyd1996}.  A rich and complex subclass of such systems are atomic nuclei, whose constituents are protons and neutrons, known together as nucleons.  
A comprehensive solution of the many-nucleon problem %
remains an outstanding challenge. 
In particular, the vast majority of dynamical processes -- such as nuclear reactions --
for the most part remains out of reach even in the age of exascale classical computing. 

Nascent demonstrations using a minimal discrete gate set~\cite{barends2015} on superconducting quantum devices have shown promise for simulating quantum systems \cite{chen2014,las2014,mezzacapo2014,roushan2014,geller2015,chiesa2015,o2016,salathe2015,neill2016,roushan2017,roushan2017spectroscopic,kandala2017,langford2017,sameti2017,dumitrescu2018,potovcnik2018,viyuela2018,kandala2019}.  
However, limitations in gate error rates and quantum-device noise undermine their efficacy when simulating real-time (unitary) evolution~\cite{klco2018}. 
Because of this,   the solution of few-nucleon problems on presently available quantum computing resources \cite{dumitrescu2018} %
has been limited to studies based on variational quantum eigensolver methods \cite{peruzzo2014} making use of schematic nuclear interaction models. 
The development of alternative, noise-resilient protocols capable of producing an efficient mapping into the quantum hardware of the interactions of microscopic systems is therefore desirable to arrive at a faithful representation of real-time many-body dynamics. 

Single-qubit gates obtained by including information about the full (multi-level) Hamiltonian of the quantum hardware are well known %
to demonstrate high-fidelity operations in superconducting circuits \cite{chow2010}.  
Multi-level superconducting devices have also been used to demonstrate sophisticated encodings with numerically optimized pulse sequences that have proven to be quite promising in the field of hardware-efficient quantum error correction \cite{ofek2016,heeres2017,hu2019}. 
In this paper, we use these insights into high-fidelity, hardware-efficient quantum computation to propose a quantum simulation of real-time nucleon-nucleon dynamics, where the propagation of the system is enacted by a single dense multi-level gate derived from the nuclear interaction at leading order (LO) of chiral effective field theory (EFT)~\cite{Epelbaum:2008ga,Machleidt:2011zz}.
This interaction displays the main features of the nuclear force, including the characteristic tensor component of the single-pion exchange potential.  

To implement the quantum simulation, we map the nuclear Hamiltonian onto a four-level superconducting circuit, specifically a three-dimensional (3D) transmon architecture \cite{paik2011}.
We enact the %
two-nucleon gate with an effective drive computed using the gradient ascent pulse engineering (GRAPE) \cite{khaneja2005} algorithm.  %
Using the open source quantum optics toolbox (QuTIP) \cite{johansson2013}, we then simulate the output of the quantum device %
in the presence of realistic quantum hardware noise.  
We show that the simulated time-dependent probability density is only slightly attenuated as a result of the noise, thus revealing all pairwise eigenenergy differences as peaks in the spectra obtained from its  discrete Fourier transform.  We further demonstrate that, by propagating the third power of the nuclear Hamiltonian, we can extract the absolute energy eigenvalues of the simulated quantum system without the use of quantum phase estimation. 

We structure this paper as follows: In Sec.~\ref{nuclear} we describe the neutron-neutron interaction. A review of the necessary circuit quantum electrodynamics needed to implement the nuclear simulation is presented in Sec.~\ref{cqed}.  In Sec.~\ref{encoding}, we describe the mapping used to encode the nuclear degrees of freedom into a single multi-level quantum device.  
Finally, in Sec.~\ref{results} we describe quantum device-level simulations from a Lindblad master equation with realistic system noise and conclude in Sec.~\ref{conclusion}.

\section{\label{nuclear} Simulations of Nuclear Dynamics}
\begin{figure}
\includegraphics[width=80mm, angle = 0]{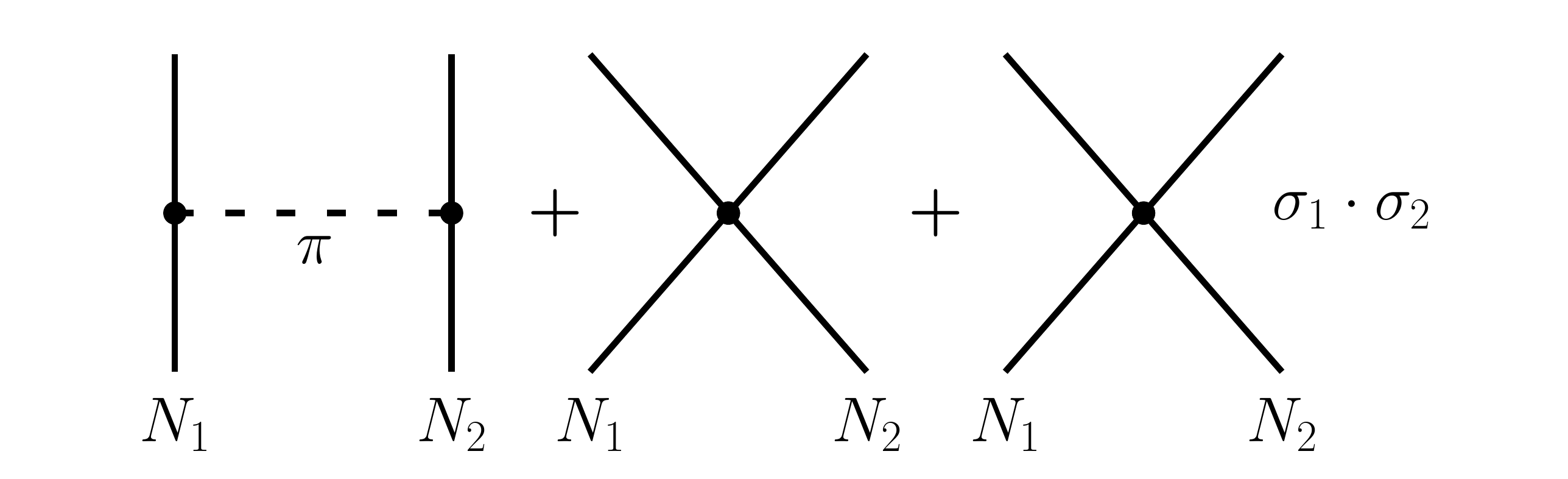}
\caption{%
Schematic description of the leading order nucleon-nucleon interaction. The left diagram depicts a single pion exchange while the middle and right diagrams depict a spin-independent and spin-dependent contact term, respectively.  %
\label{fig:chiralLO}
}
\end{figure}

 In modern nuclear theory, the description of nuclear properties and nuclear dynamics relies on an effective picture where the underlying theory of quantum chromodynamics (QCD) is translated into a systematically improvable expansion of the interactions between constituent nucleons by means of chiral EFT~\cite{Epelbaum:2008ga,Machleidt:2011zz}. 
The resulting nuclear force presents a non-trivial dependence on the spins of the nucleon pair. This dependence is manifest in two-nucleon systems, of which only the proton-neutron pair forms a bound state - the nucleus of deuterium or $^2$H - while both the proton-proton and neutron-neutron pairs are unbound.
At the same time, it was empirically recognized from an early stage that the force between two nucleons includes  a tensor-like, spin-dependent component~\cite{goodman1980,love1981,franey1985}.  The main interaction mechanism at medium distance ($\sim 2\times 10^{-15}$m) \cite{WEINBERG1990288} is the exchange of a single pion, 
while at shorter distances one can effectively recombine all the remaining processes (corresponding to the exchange of multiple pions or heavier mesons) into a contact force, depending  on the relative spin state of the nucleons. These characteristic features of the nucleon-nucleon interaction are already captured by the leading order (LO) in the chiral EFT expansion (see Fig.~\ref{fig:chiralLO}), where the Hamiltonian $\hat H^{\rm LO}$ is given by the sum of two terms:  A spin-independent (SI) component  
$\hat H_{\rm SI} = \hat T +\hat V_{\rm SI}$, where $\hat T$ is the kinetic energy of the nucleons and $\hat V_{\rm SI}$ a spin-independent portion of the two-nucleon potential; and a spin-dependent (SD) component of the interaction, $\hat V_{\rm SD}$, %
acting on the spin degrees of freedom.

In this paper, we devise a real-time propagation scheme for the quantum simulation of two interacting nucleons. 
The evolution with time $t$ of a generic state of the system $\vert\Psi\rangle$ is given by
the formal solution of the time-dependent Schr\"odinger equation for a time-independent Hamiltonian
\begin{equation}
\begin{array}{rcl}
\vert\Psi(t)\rangle &=& \exp\left[-i\hat{H}^{\rm LO}t/\hbar\right]\vert \Psi\rangle\\ \\
& =& \exp\left[-i\left(\hat{T}+\hat{V}_{\rm SI}+\hat{V}_{\rm SD}\right)t/\hbar\right]\vert\Psi\rangle\,,
\label{eq:propagation}
\end{array}
\end{equation}
where $i=\sqrt{-1}$ and $\hbar$ is the reduced Planck constant.
In the spirit of Feynman's path integrals, the propagation time can be broken up in a number of small intervals $\delta t$, and Eq.~\eqref{eq:propagation} can be well approximated by
\begin{align}
&\exp\left[-i\hat{H}^{\rm LO}\delta t\right]\nonumber\\
&\qquad\simeq \exp\left[-i\hat{H}_{\rm SI}\delta t/\hbar\right]\exp\left[-i\hat{V}_{\rm SD}\delta t/\hbar\right].
\label{eq:shorttprop}
\end{align}

\begin{figure}
\includegraphics[width=\columnwidth]{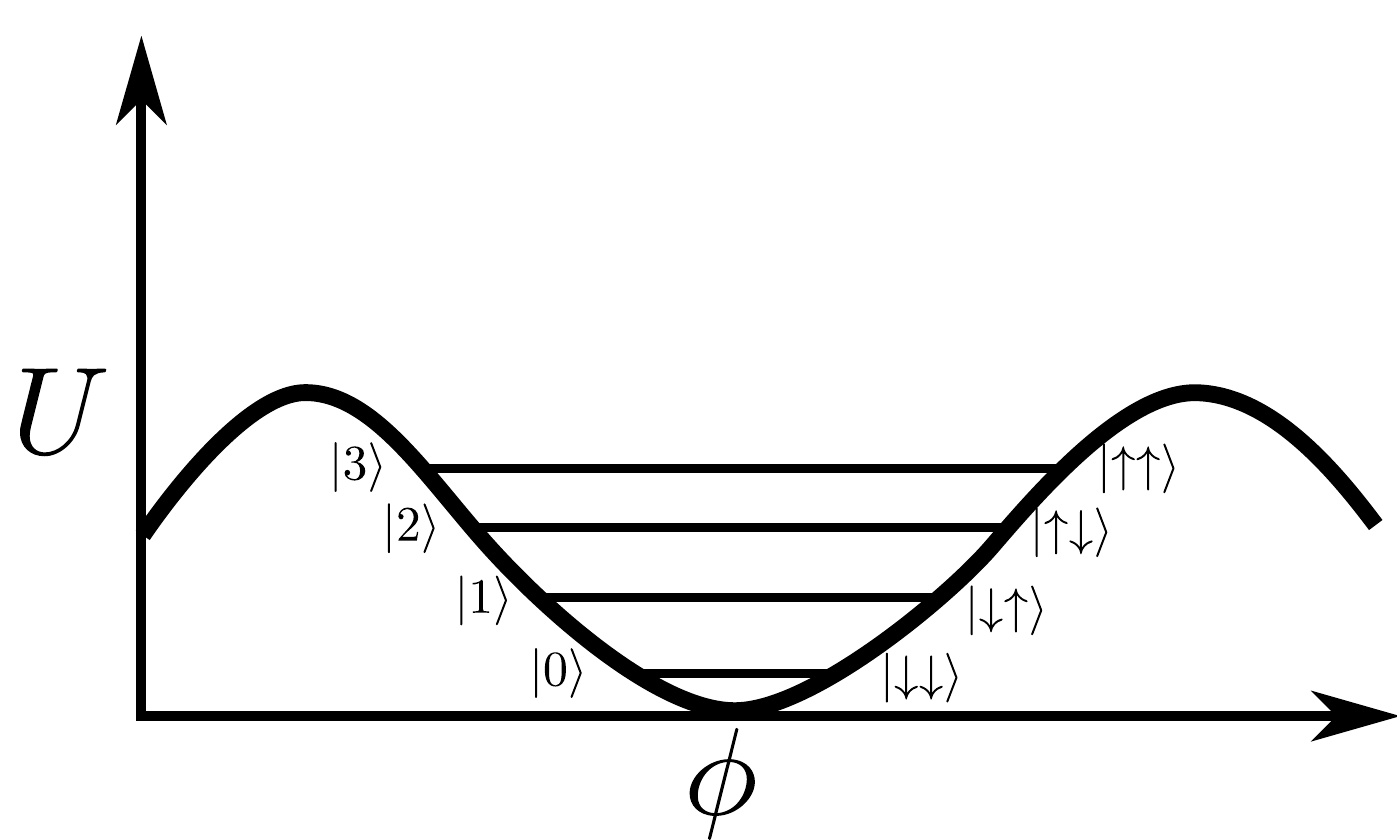}
\caption{Schematic potential energy diagram of a transmon superconducting quantum device as a function of the change of the magnetic flux ($\phi$) of the Josephson junction.  The first four energy levels are labeled in terms of their Fock number on the left side of the potential energy well while the right side shows the correspondence to two-neutron %
spin states in the independent spin basis.  
}
\label{fig:cqed}
\end{figure}

More explicitly, considering a system of two neutrons,  
the SD interaction at a separation $\vec{r}=\vec{r_1}-\vec{r_2}$  %
can be divided into a scalar and a tensor component as 
\begin{equation}\label{eq:vsd}
	\hat{V}_{\rm SD}(\vec{r}) = A^{(1)}(\vec{r}) \,\sum_\alpha \sigma^{1}_\alpha \sigma^{2}_\alpha \,+\,\sum_{\alpha,\beta} \sigma^{1}_\alpha \,  A^{(2)}_{\alpha\beta}(\vec{r}) \, \sigma^{2}_\beta\,,
\end{equation}
where $\sigma_\alpha^k$, $\alpha=x,y,z$ are Pauli matrices acting on the spin of nucleon $k=1,2$, and the functions $A^{(1)}(\vec{r})$ and $A^{(2)}_{\alpha\beta}(\vec{r})$ can be obtained from the nucleon-nucleon interaction at LO of chiral EFT in coordinate space. 
While the detailed expressions of  $A^{(1)}(\vec{r})$ and $A^{(2)}_{\alpha\beta}(\vec{r})$ bear little relevance for the present general discussion, their explicit functional form can be readily obtained from, e.g., Refs.~\cite{Gezerlis:2014} or \cite{Tews2016} and we provide an example in Appendix~\ref{Afunctions}.
 
A further approximation of Eq.~\eqref{eq:shorttprop} can be obtained by treating the neutrons as `frozen' in space 
for the duration of the spin-dependent part of the propagation, reducing
the two-neutron problem to the description of two spins
interacting through the nuclear Hamiltonian of Eq.~\eqref{eq:vsd} at a fixed seperation.
Under this approximation, the SI and SD components of the propagator in Eq.~\eqref{eq:shorttprop} act exclusively on the spatial and spin parts of the system, respectively. By projecting the state $|\Psi(t)\rangle$ onto a complete set of states $|\vec{r},s_1s_2\rangle \equiv |\vec{r}\rangle\otimes |s_1s_2\rangle$, normalized as $\langle \vec r ,s_1 s_2|\vec{x}, s^\prime_1 s^\prime_2\rangle= \delta(\vec{r}-\vec{x})\delta_{s^\prime_1 s_1}\delta_{s^\prime_2 s_2}$, the wave function at an evolved time $t+\delta t$ can be written as
\begin{align}
\label{eq:sh_time_prop}
&\langle \vec r, s_1 s_2 | \Psi(t+\delta t)\rangle\nonumber\\ 
& \simeq \!\! \displaystyle\sum_{s^\prime_1 s^\prime_2} \!\int \!\!d^3\vec x \left\langle \vec r\left| \exp[-i (\hat T + \hat V_{\rm SI})\delta t/\hbar] \right| \vec x\right\rangle \\
\nonumber& \; \times \left\langle  s_1 s_2\left| \exp[-i \hat V_{\rm SD}(\vec{x})  \delta t/\hbar] \right| s^\prime_1 s^\prime_2\right\rangle \left.\left\langle \vec{x}, s^\prime_1 s^\prime_2 \right | \Psi(t) \right\rangle.
\end{align}  
That is, for an infinitesimal time step, one can first carry out the propagation of the spin part of the wave function keeping the position of the neutrons fixed  
using only the SD part of the interaction, and then perform the spatial propagation through the SI component of the Hamiltonian. 
The present framework opens  the possibility for a classical-quantum co-processing protocol in which the propagation of the spin states is carried out by a quantum processor. For the time being we focus on the propagation of the spin-component of the two-nucleon system, that is on the application of 
the short-time propagator $\left\langle s_1 s_2\left| \exp[-i \hat V_{\rm SD}(\vec{x}) \delta t/\hbar] \right| s^\prime_1 s^\prime_2\right\rangle$. Specifically, we are interested in obtaining the probability of the occupation of each of the four possible spin states at a time $t$, starting from an initial state $|s^{(0)}_1 s^{(0)}_2\rangle$, that is 
\begin{align}
\label{eq:probEv}
P_{s_1s_2}(\vec{x},t) = \left| \left\langle s_1 s_2\left| \exp[-i \hat V_{\rm SD}(\vec{x}) t/\hbar] \right| s^{(0)}_1 s^{(0)}_2\right\rangle \right|^2.
\end{align}

 \section{\label{cqed}Circuit Quantum Electrodynamics}
We implement the %
propagator of Eq.~\eqref{eq:probEv}%
by means of a superconducting circuit quantum electrodynamic (cQED) system \cite{blais2004}.  In a cQED system, a nonlinear circuit such as a transmon \cite{koch2007,schreier2008}, plays the role of an atom coupled to the resonant mode of a microwave cavity.  In the strong, dispersive regime the resonance frequencies of each mode are separated by many line-widths as well as produce well-resolved single photon frequency shifts \cite{wallraff2004}.  In particular, we adopt a 3D transmon architecture \cite{paik2011}%
, since its long coherence times \cite{rigetti2012} and nonlinearities make it amenable to numerically optimized pulse sequences.  The full Hamiltonian for a 3D transmon coupled to a readout cavity is \cite{nigg2012}: 
\begin{equation}\label{eq:fullh}
\hat{H}_{d} =\hbar \omega_T\hat{a}_T^{\dagger}\hat{a}_T +\hbar \omega_R\hat{a}_R^{\dagger}\hat{a}_R 
- E_J\left[\cos (\hat{\phi} )+\frac{\hat{\phi}^2}{2}\right]
\end{equation}
where $\omega_{T(R)}$ and $\hat{a}^\dagger_{T(R)} (\hat{a}_{T(R)})$ are respectively the bare frequency and creation (annihilation) operators of the transmon (readout), $E_J$ is the Josephson energy, and $\hat{\phi}$ is the phase across the junction. The phase operator is given by the sum of the operators for each mode according to $\hat{\phi}=\sum_j \Phi_{zpf,j}\left(\hat{a}_j^{\dagger}+\hat{a}_j\right)$, where $\Phi_{zpf,j}$ and   $\hat{a}_j^{\dagger}(\hat{a}_j)$ are the zero-point fluctuations and the creation (annihilation) operators of the  $j^{\textrm{th}}$ mode, respectively.  A schematic of the potential energy of the transmon mode in terms of the flux is presented in Fig.~\ref{fig:cqed}. Also shown in the figure is a schematic of the first four energy levels of the transmon (labelled in terms of their Fock number), which span the computational space for the nuclear simulation of this paper.  

We make a unitary transformation into the frames of both the transmon and readout cavities 
to simplify the numerical optimization as well as to clarify the relevant quantum hardware interaction terms.  Expanding the cosine to fourth order 
we get 
\begin{eqnarray}\label{eq:simpleh}
\hat{H}^{(4)}_{d} =-\hbar\frac{\alpha_T}{2} \hat{a}_T^{\dagger2}\hat{a}_T^2 -\hbar \frac{\alpha_R}{2}\hat{a}_R^{\dagger2}\hat{a}_R^2\nonumber \\ -\hbar \chi \hat{a}_T^{\dagger}\hat{a}_T\hat{a}_R^{\dagger}\hat{a}_R + O(\hat{\phi}^6),
\end{eqnarray}
where $\alpha_{T(R)}$ corresponds to the anharmonicity of the transmon (readout) and 
$\chi$ is the dispersive interaction between the transmon and readout.  The dispersive interaction enables a quantum non-demolition readout \cite{blais2004} that -- when coupled to a phase preserving quantum limited amplifier, such as a traveling wave parametric amplifier \cite{macklin2015} -- enables high-fidelity, single-shot discrimination for all four computational states \cite{ofek2014}.

\section{\label{encoding}Hardware Efficient Encoding}
\begin{figure}
\includegraphics{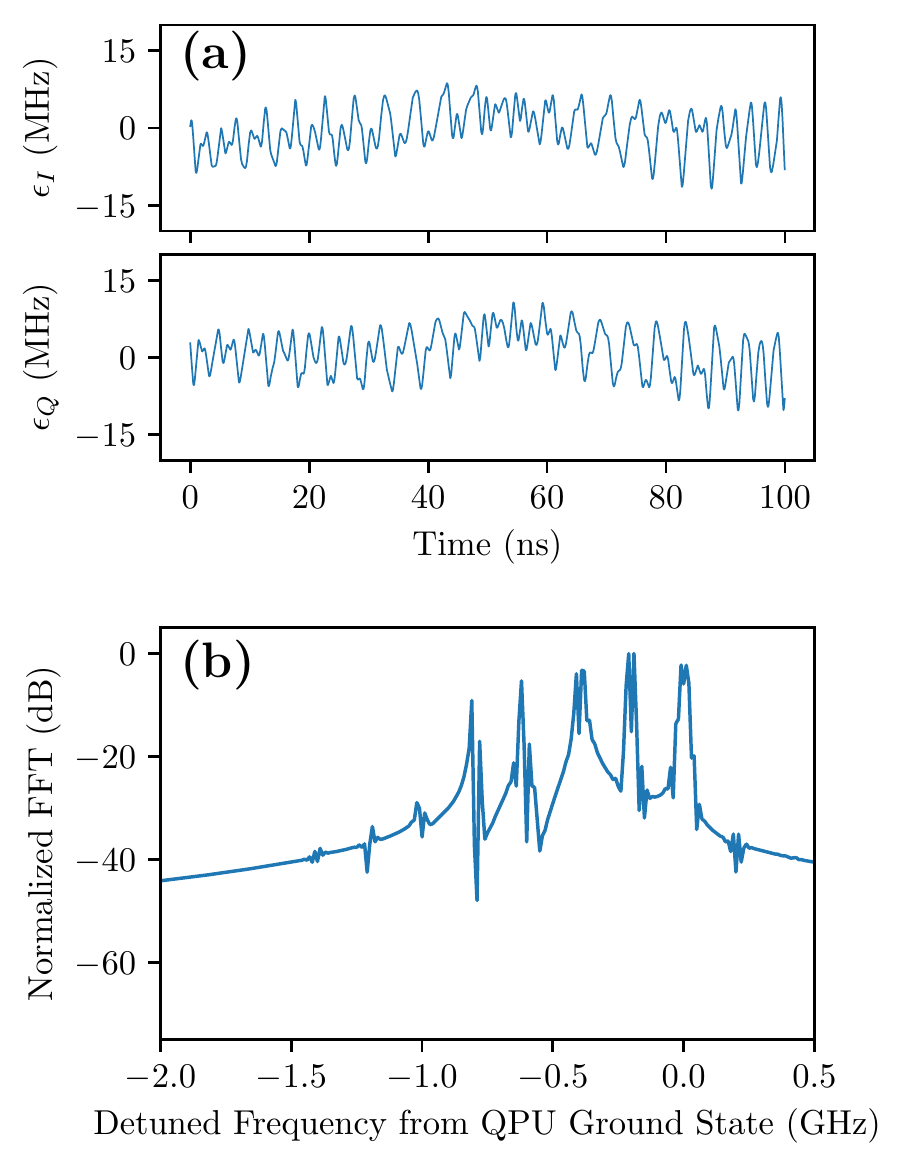}
\caption{%
\textbf{(a)} Numerically optimized time dependent drive in the rotating frame of the ground state energy transition of the transmon device that at its completion enacts the short time propagator of the nuclear Hamiltonian.  \textbf{(b)}  Discrete Fourier transform of \textbf{(a)} showing the spectral components correspond to the underlying energy transitions of the quantum device.}
\label{fig:pulse}
\end{figure}
As shown in Fig.~\ref{fig:cqed}, we use the lowest four energy levels of our superconducting quantum device to encode the spin-dependent interaction between two neutrons.  The processor mapping is as follows: The $\ket{0}$ Fock state of our processor corresponds to the uncoupled spin state of $\ket{\downarrow \downarrow}$. Likewise, $\ket{1}$, $\ket{2}$, and $\ket{3}$ correspond, respectively, to the uncoupled spin states $\ket{\downarrow\uparrow}$, $\ket{\uparrow\downarrow}$, and $\ket{\uparrow \uparrow}$.  
To implement
a single time step of the digital-time simulation %
we drive the transmon with a customized control pulse sequence. The approach adopted to obtain the optimal control is described in the following.

For a single-mode transmon we can fully describe a time-dependent drive in the frame of the transmon as \cite{heeres2017}
\begin{eqnarray}\label{eq:ctrl}
\hat{H}_c = \hbar\epsilon_{I}(t)\left(\hat{a}_T^{\dagger}+\hat{a}_T\right)+i\hbar\epsilon_{Q}(t)\left(\hat{a}^{\dagger}-\hat{a}\right),
\end{eqnarray}
where $\hat a^\dagger$ ($\hat a$) creates (destroys) an excitation in the mode and $\epsilon_{R(I)}(t)$ is a real (imaginary) time-dependent coefficient. %
For a given digital-time step $\nts$ we can then use numerical optimization to find a particular control sequence $\hat{H}_c(\tau ')$ that satisfies, within an acceptable error, the equality
\begin{align}
\label{eq:constraint}
  \exp\!\left(-\frac{i}{\hbar}\hat{V}_{SD}\nts\right) \simeq \mathcal{T}\!\exp\!\left\{-\frac{i}{\hbar}\int_0^\tau\!\! \left[\hat{H}^{(4)}_{d}+\hat{H}_c(\tau')\right]d\tau'\right\},
\end{align}
where the left-hand side of the equation corresponds to the desired short-time nuclear propagator (with the infinitesimal time step $\delta t$ now replaced by the larger, finite $\nts$).  On the right hand side of Eq.~\eqref{eq:constraint}, the notation $\mathcal{T}\!\exp$ stands for a time-ordered exponential and $\tau$ is the duration of the control pulse.  
 
We solve the numerical optimization problem of Eq.~\eqref{eq:constraint} using QuTIP. Specifically, we employ the built-in \emph{propagator} function to create the short-time propagator on the left-hand side of the equation, which then becomes the target unitary matrix for optimization using the \emph{optimize\_pulse\_unitary} function. We note that there are many other ways to create the necessary short-time propagator through matrix exponentiation \cite{moler2003}. We sample the pulse sequence at 32 gigasamples per second as we seek to leverage wide-band control electronics that have shown great promise in cQED systems \cite{raftery2017}.  
We specifically choose the pulse duration to be 100ns, which is relatively short comparing to the coherence time~\cite{Paik:2011,Rigetti:2012,Chen:2014} of a superconducting qubit, to minimize decoherence of the quantum states during the drive. We also set the maximum drive strength to be 20MHz (corresponding to a 50ns Rabi period), which can be attained in experiments for various designs of superconducting qubits~\cite{Paik:2011,Rigetti:2012,Chen:2014}.
To minimize numerical artifacts we use six levels in the 3D transmon during numerical optimization.  We find that, due to the transmon's large anharmonicity, increasing or decreasing the number of levels used in the optimization has a negligible effect on the resulting output control sequence.  The initial guess control sequence is a small ($< 2$ MHz) amplitude Gaussian drive for a duration of $100$ns.
Given the complexity of the pulse sequence required by the nuclear Hamiltonian, the optimization requires %
about 100 iterations to complete execution with an infidelity threshold of less than $10^{-4}$. 

A typical result for the amplitudes of the control coefficients entering in Eq.~\eqref{eq:ctrl}
is shown in Fig.~\ref{fig:pulse}a. The discrete Fourier transform of this amplitude (shown in Fig.~\ref{fig:pulse}b)  highlights the underlying spectral features of the drive. One can recognize peaks corresponding to the transitions between states of the transmon. That is, the optimization procedure finds the best time filter to enact the desired nuclear Hamiltonian by driving the different energy transitions of the 3D transmon. Regardless of 
the initial conditions, when driven with this control pulse, the system does not experience state leakage out of the computational manifold.  Shorter pulse sequences (smaller $\tau$) are possible but require larger amplitude drives and the duration of the control pulse sequence is strongly dependent on the maximally accepted drive amplitude consistent with Ref. \cite{leung2017}.

\section{\label{results}Simulated Output and Validation of the Quantum Device}
\begin{figure*}
\includegraphics[width=\textwidth]{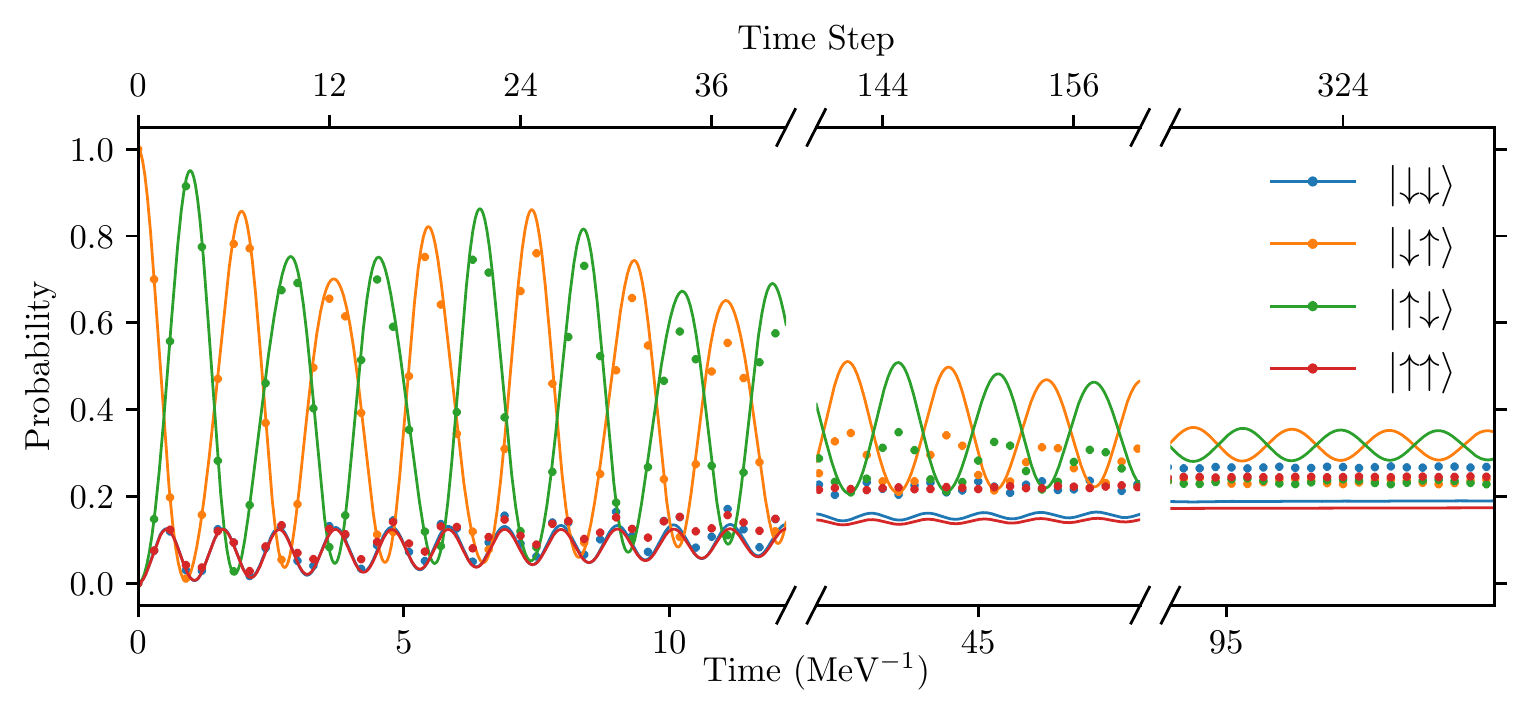}

\caption{Occupation probabilities as a function of time.  Colored circles depict the output probability as a function of simulation time step.  Each point is obtained by solving the Lindblad master equation and determining the overlap with the particular nuclear spin (Fock state) of interest.  The simulation includes dissipation and dephasing terms common in 3D transmon systems.  Solid lines result from direct integration of the interaction Hamiltonian and use dissipation and dephasing quantum device terms appropriately scaled to nuclear interaction strengths. The collapse of the four
probabilities at later times is a result of device decoherence.}
\label{fig:qp}
\end{figure*}

We investigate the performance of our numerically optimized pulse sequences by using a Markovian Lindblad master equation, 
\begin{gather}\label{eq:me}
\frac{\partial\rho}{\partial \tau}= -\frac{i}{\hbar}[\hat{H}_{\rm QPU},\rho]+\left(\frac{1}{T_1}D[\hat{a}]+\frac{1}{T_{\phi}}D[\hat{a}^{\dagger}\hat{a}]\right)\rho\\
\nonumber\hat{H}_{\rm QPU} = \hat{H}_d^{(4)} + \hat{H}_c(\tau)\\
\nonumber D[\hat{a}]\rho= \hat{a}\rho\hat{a}-\frac{1}{2}\{\hat{a}^{\dagger}\hat{a},\rho\}\,,
\end{gather}
which is well suited for modeling the density matrix $\rho$ of driven dissipative cQED systems \cite{murch2012,geerlings2013,leghtas2013,holland2015,leghtas2015,heeres2017}.
Here,  $T_1$ is the transmon energy relaxation time, which, for these simulations, we have assumed to be $30~\mu s$, and $T_\phi$ is the transmon dephasing time, which is taken to be $50~\mu s$. This yields a total coherence time of $T_2^*\approx27~\mu s$, shorter than the state of the art \cite{heeres2017}, giving us a conservative estimate of the efficacy of our approach.  Furthermore, we have assumed a typical 3D transmon anharmonicity value of $\alpha_T=200$ MHz.

In Fig.~\ref{fig:qp}, we present (as solid lines and circles) the time-dependent occupation probabilities of the two-neutron spin states obtained from two different Lindblad master-equation simulations.
Specifically, the solid lines depict the solution obtained from propagating the neutrons' spin-states using the exact spin-dependent term of the nuclear Hamiltonian [Eq.~\eqref{eq:vsd}] with device noise terms scaled to the relevant nuclear interaction strengths.  The circles represent the simulated output probability distributions from the quantum device at the culmination of a single pulse sequence (the intermediary behavior during the application of the real-time propagation gate is not shown) obtained with repeated applications of the control sequence.  
As time progresses, the
quantum device -- initially
prepared in the $\ket{\downarrow\uparrow}$ state --
evolves into an entangled
superposition of the four spin
states. More interestingly, we
can observe multiple entire
cycles of the dynamics before
the device reaches decoherence.

In the following, we show that the time dependence of the occupation probabilities for each state encodes
the eigenvalues of the spin-dependent term of the nuclear Hamiltonian ($V_{\rm SD}$). 
\begin{figure}
\includegraphics{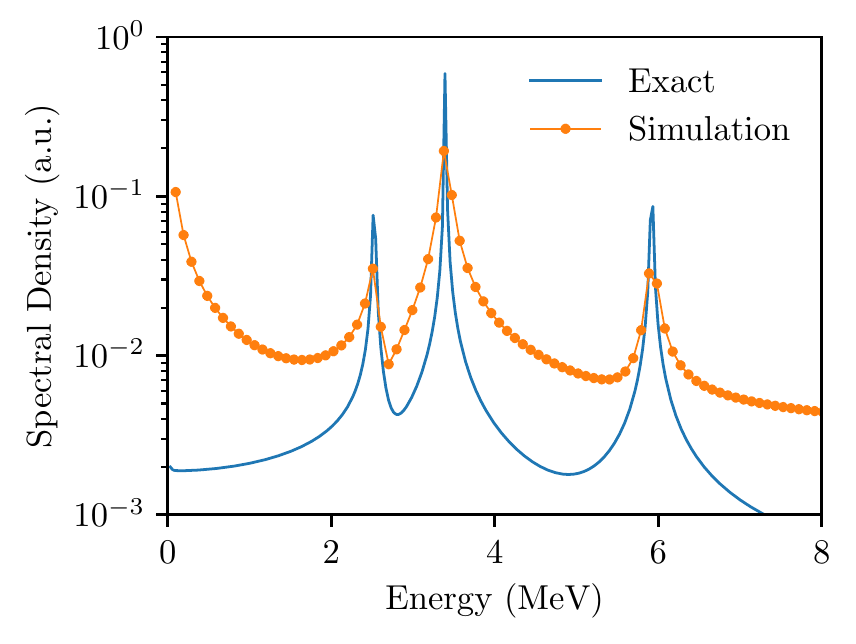}
\caption{Energy Spectra. The solid line is the Fourier transform of the $\ket{\downarrow\uparrow}$ state (first Fock state) time-dependent probability distribution for the exact n-n interaction propagator (without noise).  The spectral lines correspond to all possible eigenenergy differences.  The circles correspond to the Discrete Fourier Transform of the $\ket{\downarrow\uparrow}$ state (first Fock state) time dependent probability distribution for repeated application of the n-n interaction digital time propagator (including realistic 3D transmon decoherence).  Even with decoherence, the pairwise eigenenergies differences are discernible in the spectra.}
\label{fig:fft}
\end{figure}
In general, any state $|\Psi(t)\rangle$ that results from the application of the nuclear %
propagator can be decomposed into the basis defined by the eigenvectors $\ket{\phi_j}$ of the corresponding driving Hamiltonian (in this case $\hat{V}_{\rm SD} \ket{\phi_j} = \lambda_j \ket{\phi_j}$). Since the latter is time-independent, the expansion takes the form
\begin{equation}
\left| \Psi(t) \right\rangle = \displaystyle\sum_j c_j e^{-i\hat{V}_{\rm SD} t/\hbar} \left| \phi_j \right\rangle = \displaystyle\sum_j c_j e^{-i \lambda_j t/\hbar} \left| \phi_j\right\rangle, 
\end{equation}
 with coefficients $c_j = \langle \phi_j | \Psi(t) \rangle$. Introducing the basis $|\xi_i\rangle$, on which device measurements are made, we can readily obtain an expression for the time dependence of the probability of measuring each state 
\begin{align}
\nonumber \left| \langle\xi_i | \Psi(t) \rangle \right|^2 &= \left| \displaystyle\sum_{j} c_j b^j_m e^{-i \lambda_j t/\hbar} \right|^2 \\ &=  \displaystyle\sum_{jk} c_j b_m^j c_k^* {b_m^k}^* e^{-i\Delta \lambda_{jk}t/\hbar},
\label{eq:ProbabilityEvolution}
\end{align}
where we have introduced the notation $\Delta \lambda_{jk} = \lambda_j - \lambda_k$ for the difference between any pair of eigenvalues and the overlap $b_j^i= \langle \xi_i | \phi_j \rangle$. The consistency of the eigenvalue differences extracted from the device's signal with those computed analytically can be used as a validation of the encoding of the nuclear Hamiltonian and the quantum simulation of its time evolution. This can be readily achieved by analyzing the Fourier transform of the occupation probabilities.

In Fig.~\ref{fig:fft}, we display the sum of the squared magnitudes of the discrete Fourier transforms of the simulated occupation probabilities (i.e., the power spectra) obtained with the system prepared at $t=0$ in the state $\ket{\downarrow\uparrow}$ (which has non-zero overlap with all the eigenstates of the spin-dependent interaction Hamiltonian).   The solid line with circles was computed using the solution of the master equation (\ref{eq:me}), whereas the plain solid line (without circles) is the corresponding result obtained %
from the evolution of the exact Hamiltonian in the absence of noise. In a system spanning $d$ different states, as long as the state $|\Psi(t)\rangle$ has some non-zero overlap with all measurement states, we expect to see $d(d-1)/2$ peaks in the power spectra. 
The degeneracy in two of the eigenvalues reduces the total number of peaks seen in the spectrum in Fig.~\ref{fig:fft} to three (see Appendix~\ref{Afunctions}), corresponding to all the distinct pairwise differences of the four eigenvalues of $\hat{V_{\rm SD}}$.
Comparing the locations of these spectral peaks ($\omega_{i}$) with the $\Delta\lambda_{jk}$ provides a first validation 
of the quantum simulation.

The locations of the peaks in the discrete power spectra yield a good initial estimate of 
the physical values of 
$\omega_{i}$. However, in general such values are not contained in the discrete set of Fourier frequencies.  Therefore, we adjust our estimates by fitting both power-spectra and probabilities against their exact analytical forms combined with correlated Gaussian noise in the time domain described by a given covariance matrix $\bm{\Sigma}$. The details of this fitting procedure %
are described in Appendix~\ref{FourierA}. 
The results of this analysis are summarized in Table \ref{Tab:differences}, where we compare the extracted $\omega_{i}$ values and exact pairwise differences $\Delta\lambda_{jk}$ between the four eigenvalues of $\hat{V}_{\rm SD}$.   
\begin{table}
\caption{\label{Tab:differences}Summary of the differences $\Delta\lambda_{jk}$ of the eigenvalues of the operator $\hat V_{\rm SD}$ computed analytically and simulated with a noisy time-propagation of the spin states, as described in text.
The errors have been attributed using the analysis described in detail in Appendix~\ref{FourierA}}
\begin{ruledtabular}
\begin{tabular}{ccccc}
$\Delta\lambda_{jk}$&Exact (MeV)& $\omega_i$&Simulated (MeV) %
\\
\hline
$\Delta \lambda_{21}$ & 2.5254 & $\omega_2$ & 2.55(2) 	%
\\
$\Delta \lambda_{10}$ & 3.3951 & $\omega_4$ & 3.41(3) 	%
\\
$\Delta \lambda_{20}$ & 5.9205 & $\omega_5$ & 5.93(1)	%
\\
\end{tabular}
\end{ruledtabular}
\end{table}
As it can be seen, the %
two sets of values are in fair agreement. The error of order $10^{-2}$ associated with the extraction of the physical values $\omega_i$ provides an estimate of the resolution of our quantum simulation. 

In the remainder of this section, we show that the absolute eigenvalues $\lambda_k$ can be further extracted
without the use of quantum phase estimation. Towards this aim, we label the eigenvalues from smallest ($\lambda_0$) to largest ($\lambda_3$).  
Out of the three $\Delta \lambda_{jk}$ combinations shown in Table \ref{Tab:differences}, the largest value corresponds to the difference  $ \lambda_3-\lambda_0 = \alpha$. 
By carrying out a second quantum simulation in which the real-time propagation of the system is driven by the third power of the nuclear interaction Hamiltonian, $\hat{V}_{SD}^3$, and again Fourier transforming the output probability distributions, we can then find a second set of eigenvalue differences. %
Such a propagation has the advantage of leaving the resulting eigenvectors unchanged but yielding new eigenvalues $\lambda_k^3$. The locations of the new peaks are now found at the differences of the cubes of the eigenvalues,
with the largest value corresponding to $\lambda_3^3-\lambda_0^3=\beta$.
As a result, we obtain a non-linear system of two equations with two unknown parameters, which presents two possible solutions for the extremal pair of eigenvalues ($\lambda_0,\lambda_3$)
\begin{align}
\lambda_0 &= -\frac{\alpha}{2} \pm \sqrt{\frac{\beta}{3\alpha} - \frac{\alpha^2}{12}}\\
\lambda_3 & = \alpha + \lambda_0.
\end{align}
Of the two remaining eigenvalues $\lambda_1, \lambda_2$ one will be degenerate, with three possible cases, $\lambda_1 = \lambda_0, \lambda_1=\lambda_2$ or $\lambda_2=\lambda_3$, yielding a total of 6 distinct combinations of eigenvalues. For each combination, the non-degenerate $\lambda_{1(2)}$ eigenvalue is obtained by solving for the trace trace of the nuclear interaction Hamiltonian. Because in the specific case considered here this matrix is traceless, we modify it by adding a constant diagonal matrix so  to induce a non-zero value for the trace. 
We note that even though  in principle the dimensions of the Hamiltonian matrix could be large, the trace still scales linearly with increasing matrix size. 
Only one out of the six possible combinations of eigenvalues  will closely reproduce the set of $\Delta\lambda_{jk}$, and this criterion is used to determine the correct eigenvalues.
The absolute eigenvalues %
obtained from this analysis are summarized in Table \ref{Tab:eigen}. For the extremal eigenvalues, we find overall good agreement between the exact values and those obtained from the noisy simulation of the system evolution. The error arising from the estimation of the peak positions compounds for the case of $\lambda_2$. 
\begin{table}
\caption{\label{Tab:eigen}Summary of the eigenvalues of the
operator $\hat V_{\rm SD}$ from exact analytic calculations and as extracted from the real-time nuclear simulation. Errors have been obtained by propagating the uncertainties reported in Tab. \ref{Tab:differences}, as well as the ones from the propagation of $V_{\rm SD}^3$.}
\begin{ruledtabular}
\begin{tabular}{cccc}
Eigenvalue&Exact (MeV)&Simulated (MeV)\\
\hline
$\lambda_{0,1}$ & -2.329 & -2.3(2) \\
$\lambda_2$ & 1.066 & 0.9(6) \\
$\lambda_3$& 3.592 & 3.6(2) \\
\end{tabular}
\end{ruledtabular}
\end{table}

\section{\label{conclusion}Conclusion}
We presented a single-gate approach based on efficient quantum-hardware mapping for realizing the real-time evolution of the spin states of two interacting neutrons on a multi-level superconducting quantum processor. The interaction Hamiltonian for the nuclear spins is obtained from the neutron-neutron interaction at leading order of chiral effective field theory by fixing the relative position of the neutrons, and retaining only the spin-dependent components of the resulting potential.  The single, multi-level gate required for the faithful encoding of the nuclear short-time propagator onto the quantum device is obtained by numerical optimization, by leveraging the well known device Hamiltonian of 3D transmons.  

To investigate the performance of our approach, we used a Markovian Linblad master equation to model the 
output of the quantum device -- initially prepared in the $\ket{\downarrow\uparrow}$ state and then driven by the numerically-optimized 
pulse (gate)
-- in the presence of realistic quantum-hardware noise.  The resulting simulated output occupation probability shows that, with the progression of time, the quantum device evolves into an entangled superposition of the four spin states, and that the signal is only slightly attenuated as a result of the noise. 

Finally, we showed that thanks to the longevity of the signal enabled by our single-gate approach to real-time propagation, one can then compute the Fourier transform of the occupation probability and extract information about the energy spectrum of the simulated nuclear system, which is one of the fundamental properties one is interested in describing when solving any many-body problem. Specifically, we related the characteristic  peak structure of the computed power spectral density to the pairwise differences of the eigenenergies of the adopted interaction Hamiltonian for the nuclear spins. We then demonstrated, by additionally carrying out the real-time propagation of the third power of the nuclear Hamiltonian, that we can extract the absolute energy eigenvalues of the simulated quantum system without the use of quantum phase estimation. 

In the present application, we confined our study to a system of two neutrons at fixed relative position, thus disregarding the evolution of their spatial wave function. More in general, the real-time propagation scheme introduced in Sec.~\ref{nuclear} opens the possibility for a  classical-quantum co-processing protocol in which the propagation of the spin states is carried out by a quantum processor while the spatial propagation is performed with classical computing. Such a protocol would provide a pathway to addressing the exponentially growing number of spin configurations with increasing number of nucleons, which is currently a major computational bottleneck in simulating real-time evolution with quantum Monte Carlo methods. 
   
Finally, we note that the methods discussed in this paper can be readily applied to a wide range of real-time quantum simulations and are not restricted to nuclear physics problems.  Therefore, this work opens a meaningful pathway for enabling transformative quantum simulations during the noisy intermediate scale quantum hardware era.

\acknowledgments
This work was performed under the auspices of the U.S. Department of Energy by Lawrence Livermore National Laboratory under Contract DE-AC52-07NA27344.  This work was supported by the Laboratory Directed Research and Development grants 19-ERD-013 and 19-DR-005.  J.D, X.W. and E.H. acknowledge partial support by the DOE ASCR quantum testbed pathfinder program.

\appendix
\section{}
\label{Afunctions}
Following the notation of Ref.~\cite{Tews2016}, the explicit form of the functions $A^{(1)}(\vec r)$ and $A^{(2)}_{\alpha,\beta}(\vec{r})$ appearing in the expression of the SD neutron-neutron interaction at LO of chiral EFT in coordinate space [see Eq.~\eqref{eq:vsd}] are given by
\begin{align}
  A^{(1)}(\vec r) 
    &= 
      C_1\delta_{R_0}(\vec{r}) 
      - 
      Y_{\pi}(r)
      \left(1- e^{-\left(r/R_0\right)^4}\right)
  \\
  A^{(2)}_{\alpha, \beta}(\vec{r}) 
    &=
      T_{\pi}(r) 
      \left(3\frac{r_\alpha r_\beta}{r^2} - \delta_{\alpha\beta}\right) 
      \left(1- e^{-\left(r/R_0\right)^4}\right)
      \,.
\end{align}
Similarly, the spin-independent part of the interaction can be written as $\hat V_{\rm SI}(\vec r) =C_0\,\delta_{R_0}(\vec r)$.
In the above expressions,
\begin{equation}
  \delta_{R_0}(\vec r) = \frac{1}{\pi\Gamma(3/4)R_0^3}\, \exp(-r/R_{0})
\end{equation} 
is a regulated Dirac $\delta(\vec r)$ function, $C_1$ and $C_2$ are constants that are typically fitted to reproduce some experimental quantity  (such as, e.g., the $s$-wave nucleon-nucleon phase shifts), and
\begin{equation}
  Y_\pi(r) 
    = 
      \frac{m_\pi^3}{12\pi}
      \left(\frac{g_a}{2f_\pi}\right)^2
      \frac{\exp(-m_\pi r)}{m_\pi r}
\end{equation} 
and 
\begin{equation}
  T_\pi(r) = \left( 1+ \frac{3}{m_\pi r} + \frac{3}{m_\pi^2 r^2} \right) \, Y_\pi(r)\,
\end{equation}
are functions entering the definition of the one-pion exchange potential where $g_a$, $f_\pi$ and $m_\pi$ are respectively the axial-vector coupling constant, the pion decay constant, and pion mass.  

The spin eigenvalue decomposition of this Hamiltonian can be computed exactly for all values of the internuclear separation $r$, and yields 3 distinct eigenvalues: $-3 a$, $a - 4 b$, and $a + 2 b$,  where $a=A^{(1)}(\vec r)$ and $b = T_{\pi}(r) \left(1- e^{-\left(r/R_0\right)^4}\right)$. The last eigenvalue ($a + 2 b$) is associated with 2 degenerate eigenstates.

In this work, we choose $\left| \downarrow  \uparrow \right\rangle$ as the initial state of the system.  
Letting $\vec{r} = r \left(\sqrt{1-x^2}\cos \phi \hat{e}_x + \sqrt{1-x^2}\sin \phi \hat{e}_y + x \hat{e}_z\right)$, the overlaps of the initial state with the eigenstates of the nuclear interaction Hamiltonian  are
\begin{align*}
  \left\langle \downarrow \uparrow \middle | -3 a \right \rangle 
    &= 
      -\frac{1}{\sqrt{2}}
  \\
  \left\langle \downarrow \uparrow \middle | a - 3 b \right \rangle 
    &=
      -\frac{x e^{i\phi}}{\sqrt{2}}
  \\
  \left\langle \downarrow \uparrow \middle | a + 2 b \right \rangle_1
    &=
      \frac{1}{\sqrt{2}}\sqrt{\frac{1-x^2}{1+x^2}}
  \\
  \left\langle \downarrow \uparrow \middle | a + 2 b \right \rangle_2
    &=
    \frac{x e^{i\phi}}{\sqrt{2}}\sqrt{\frac{1-x^2}{1+x^2}}\,.
\end{align*}

The nuclear Hamiltonian is rotationally invariant. As a result of this, the spectra of our frozen system are independent of the direction $\hat{r}$.  Rather than trivially choosing $\hat{r}$ to lie along the z-axis,  which would have resulted in only two of the four states mixing during the evolution, here we choose  $\hat{r}$ to point in a random direction, %
allowing us to explore more general cases one may encounter  in an actual implementation of this Hamiltonian on a QPU. The results presented in this work were obtained with $x = 0.382$, $\phi = 2.71\,\rm{degrees}$ and $r=3.5\,\rm{fm}$.  We used a time step of $0.30\,\rm{MeV}^{-1}$.

\section{}
\label{FourierA}
The analysis of the uncertainty on the peak positions of the Fourier transform begins with the assumption that time-correlated Gaussian noise is sufficiently descriptive of QPU noise to give us reasonable  extraction of the peak locations.  The probability of measuring state $m$ at time $t$ is parameterized as
\begin{align}
  g_m(t, \bm{\omega}) 
    &= 
      d_{m} 
      + 
      \sum_{j} 
        d_{m, j} \cos(\bm{\omega}_j t)
        +
        \tilde{d}_{m, j} \sin(\bm{\omega}_j t)\,
\end{align}
where the set $\bm{\omega}$ are the locations of the peaks in the power spectra.  The real coefficients $d_{m}$, $d_{m,j}$, and $\tilde{d}_{m,j}$ are constrained using generalized least squares with covariance matrix $\bm{\Sigma}$, leaving the peak frequencies $\bm{\omega}$ and the parameters needed to construct $\bm{\Sigma}$ as the only free parameters to adjust.  In the exact evolution, $d_m$ can be related to the coefficients in Eq.~\ref{eq:ProbabilityEvolution}, and similar relationships exist for $d_{m,j}$, and $\tilde{d}_{m,j}$.  In the frequency domain, the power spectra at the frequencies $f_j$ of the discrete Fourier transform are
\begin{align}
  F_m (f_j, \bm{\omega}) 
    &= 
      \left| \sum_k {\bm{F}_{jk} g_m(t_j, \bm{\omega})} \right|^2
      +
      \left(
        \bm{F}^{\vphantom{\dagger}} \bm{\Sigma} \bm{F}^\dagger
      \right)_{jj},
\end{align}
where $\bm{F}_{jk}$ is the discrete Fourier transform matrix and the Gaussian noise has been marginalized over analytically.  We parametrize the covariance matrix as
\begin{align} 
  \bm{\Sigma}_{jk} 
    &= 
      \sigma^2 \kappa_{\min{j,k}}
     e^{-\frac{(j-k)^2}{2 l^2}}
\end{align} 
where $l$ is the correlation length of our time-domain ``noise'',  $\sigma$ is a parameter we fit that describes the overall scale of the system noise,
and  $\kappa_{a}$ linearly interpolates between $\kappa_{a} = 1$ to $\kappa_{n}=\gamma$ where $n$ is the number of time steps in the fit, and $\gamma$ is fitted to account for dissipation. The off-diagonal kernel, $e^{-\frac{(j-k)^2}{2 l^2}}$, is a simplistic way to account for the noise being time dependent (i.e dephasing at time $t_i$ is going to depend on state of system at earlier times, similar for infidelity).  In practice, the specific details of $\bm{\Sigma}$ have little impact on the final predictions for $\bm{\omega}$ once there are enough degrees of freedom.
Finally, we constrain the peak frequencies by maximizing the likelihood of the $F_m$'s assuming a Gaussian likelihood function with Gaussian priors on the frequencies $\bm{\omega}$.  The priors are centered around the initial peak estimates from the discrete Fourier transform with a $1\sigma$ width set by the difference between adjacent frequencies.

\bibliographystyle{apsrev4-1}

\end{document}